\documentclass[12pt,a4paper]{article}
\usepackage{amsmath}
\usepackage{bm}
\usepackage{amsfonts}
\usepackage{graphicx}
\usepackage[normal]{caption2}
\usepackage{subfigure}
\usepackage{rotating}
\usepackage{citesort}
\usepackage{lscape}
\usepackage{section}
\begin{document}
\renewcommand\arraystretch{1.1}
\setlength{\abovecaptionskip}{0.1cm}
\setlength{\belowcaptionskip}{0.5cm}
\title { Stability of fragments and study of participant-spectator matter at peak center-of-mass energy}
\author {Sukhjit Kaur\\
\it House No. 465, Sector-1B, Nasrali, Mandi Gobindgarh-147301,
Punjab, India\\} \maketitle
Electronic address:~sukhjitk85@gmail.com

We simulate the central reactions of nearly symmetric, and
asymmetric systems, for the energies at which the maximum
production of IMFs occurs (E$_{c.m.}^{peak}$).This study is
carried out by using hard EOS along with cugnon cross section and
employing MSTB method for clusterization. We study the various
properties of fragments. The stability of fragments is checked
through persistence coefficient, gain term and binding energy. The
information about the thermalization and stopping in heavy-ion
collisions is obtained via relative momentum, anisotropy ratio,
and rapidity distribution. We find that for a complete stopping of
incoming nuclei very heavy systems are required. The mass
dependence of various quantities (such as average and maximum
central density, collision dynamics as well as the time zone for
hot and dense nuclear matter) is also presented. In all cases
(i.e., average and maximum central density, collision dynamics as
well as the time zone for hot and dense nuclear matter) a power
law dependence is obtained.
 \par
\section{Introduction}
Heavy-ion collisions at intermediate energies offer an opportunity
of producing nuclear systems under the extreme conditions of
temperature and density. At high excitation energies, the
colliding nuclei compress each other as well as heat the matter
\cite{goss97,khoa92,khoaetal}. This leads to the destruction of
initial correlations, as a result the matter becomes homogeneous
and one can have global stopping. The global stopping is defined
as the randomization of one-body momentum space or memory loss of
the incoming momentum. The degree of stopping, however, may vary
drastically with incident energies, mass of colliding nuclei and
colliding geometry. The degree of stopping has also been linked
with the thermalization (equilibrium) in heavy-ion collisions.
More the initial memory of nucleons is lost, better it is stopped.
\par
The fragmentation of colliding nuclei into several pieces of
different sizes is a complex phenomenon. This may be due to the
interplay of correlations and fluctuations emerging in a
collision. Several studies, in literature, have been made to check
the fragmentation pattern. Fragmentation pattern has been found to
depend on the size of the colliding nuclei, incident energy as
well as impact parameter \cite{goss97,aich91,tsang93,will97}.
Dhawan \emph{et al}. \cite{dhawan06} studied the degree of
stopping reached in intermediate energy heavy-ion collisions. They
found that degree of stopping decreases with increase in impact
parameter as well as at very high energies. They suggested that
the light charged particles (LCPs) (2$\leq$A$\leq$4), can be used
as a barometer for studying the stopping in heavy-ion collisions.
As lighter fragments mostly originate from midrapidity region
whereas intermediate mass fragments (IMFs) originate from surface
of colliding nuclei and can be viewed as remnants of the spectator
matter. On the other hand, Sood and Puri \cite{sood06} studied the
thermalization achieved in heavy-ion collisions in terms of
participant-spectator matter. They found that
participant-spectator matter depends crucially on the collision
dynamics as well as history of the nucleons and important changes
in the momentum space occur due to the binary nucleon-nucleon
collisions experienced during the high dense phase. The collisions
push the colliding nucleons into midrapidity region responsible
for the formation of participant matter. This ultimately leads to
thermalization in heavy-ion collisions. Vermani \emph{et al}.
\cite{sood06} used rapidity distribution of nucleons to
characterize the stopping and thermalization of the nuclear
matter. They found that nearly full stopping is achieved in
heavier systems like $^{197}$Au+$^{197}$Au whereas in lighter
systems a larger fraction of particles is concentrated near target
and projectile rapidities, resulting in a broad Gaussian shape.
The lighter systems, therefore, exhibit larger transparency
effect, i.e., less stopping. Puri \emph{et al}. \cite{puri94}
studied the non-equilibrium effects and thermal properties of
heavy-ion collisions. They found that the heavier masses are found
to be equilibrated more than the lighter systems.
\par
 Recently,
Sisan $\textit{et al.}$ \cite{sisan01} studied the emission of
IMFs from central collisions of nearly symmetric systems using a
4$\pi$-array set up, where they found that the multiplicity of
IMFs shows a rise and fall with increase in the beam energy. They
observed that E$_{c.m.}^{max}$ (the energy at which the maximum
production of IMFs occurs) increases linearly with the system
mass, whereas a power-law ($\propto$ A$^{\tau}$) dependence was
reported for peak multiplicity of IMFs with power factor $\tau$ =
0.7. Though percolation calculations reported in that paper failed
to explain the data, subsequent calculations using QMD model
\cite{sukh10} successfully reproduced the data over entire mass.
One is, therefore, interested to understands how nuclear dynamics
behaves at this peak energy i.e. whether linear increase reported
for the multiplicity of fragments remains valid for other
observables or not. We here plan to investigate the degree of
stopping reached and other related phenomena in heavy-ion
reactions at peak center-of-mass energies. We also check system
size dependence of average and maximum central density, collision
dynamics as well as the time zone for hot and dense nuclear matter
at peak center-of-mass energy.
\par
This study is made within the framework of the quantum molecular
dynamics model, which is described in detail in Refs.
\cite{goss97,aich91,sukh10,rkpuri,sood09,jsingh00,kumarpuri98,jsingh,puri00,puri02}.

\section{Results and Discussion}
For the present study, we simulate the central reactions (b =0.0
fm) of $^{20}$Ne+$^{20}$Ne, $^{40}$Ar+$^{45}$Sc,
$^{58}$Ni+$^{58}$Ni, $^{86}$Kr+$^{93}$Nb, $^{129}$Xe+$^{118}$Sn,
$^{86}$Kr+$^{197}$Au and $^{197}$Au+$^{197}$Au at the incident
energies at which the maximal production of intermediate mass
fragments (IMFs) occurs. These read approximately 24, 46, 69, 77,
96, 124, and 104 AMeV, respectively for the above mentioned
systems  \cite{sukh10}. Note that in lower energy region phenomena
like fusion, fission and cluster decay are dominant \cite{dutt10}.
Here we use hard (labeled as Hard) equation of state along with
energy dependent Cugnon cross section ($\sigma$$_{nn} ^{free}$)
\cite{kumarpuri98}. The reactions are followed till 200 fm/c. The
phase-space is clusterized using minimum spanning tree method with
binding energy check (MSTB). The MSTB method is an improved
version of normal MST method \cite{jsingh}. Firstly, the simulated
phase-space is analyzed with MST method and pre-clusters are
sorted out. Each of the pre-clusters is then subjected to binding
energy check \cite{sukh10,jsingh}:
\begin{equation}
\zeta_{i} =
\frac{1}{N^{f}}\sum_{i=1}^{N^{f}}[\frac{(\textbf{p}_{i}-\textbf{P}_{N^{f}}^{c.m.})^{2}}{2m_{i}}
+ \frac{1}{2}\sum_{j \neq
i}^{N^{f}}V_{ij}(\textbf{r}_{i},\textbf{r}_{j})]< E_{bind}.
\end{equation}
We take $E_{bind}$ = -4.0 MeV if $N^{f}\geq 3$ and $E_{bind}$ =
0.0 otherwise. Here $N^{f}$ is the number of nucleons in a
fragment and $\textbf{P}_{N^{f}}^{c.m.}$ is center-of-mass
momentum of the fragment. This is known as Minimum Spanning Tree
method with Binding energy check (MSTB) \cite{sukh10,jsingh}. Note
that nucleons belong to a fragment if inequality (1) is satisfied.
The fragments formed with the MSTB method are more reliable and
stable at early stages of the reactions.
\par
One of the important aspects in fragmentation is the stability of
fragments as well as surrounding nucleons of a fragment. The
change in the nucleon content of fragments between two successive
time steps can be quantified with the help of persistence
coefficient \cite{puri00,puri02}.
\begin{figure}[!t]
\centering
 \vskip -1cm
\includegraphics[angle=0,width=15cm]{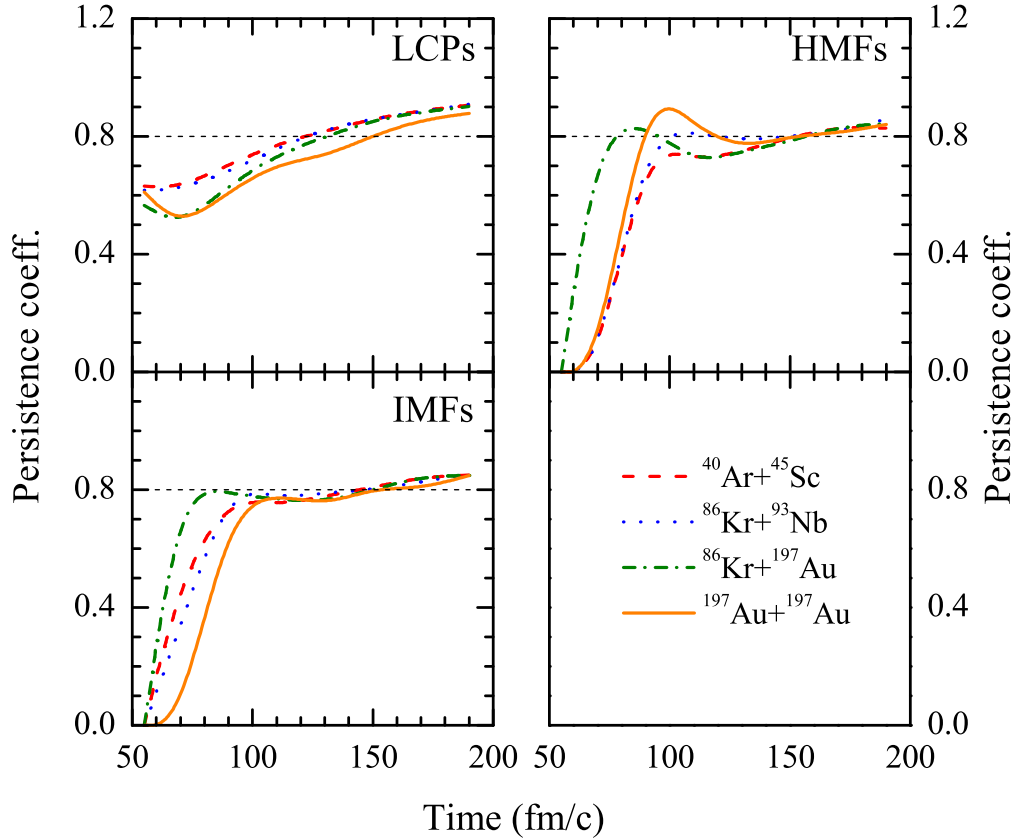}
 \vskip -6cm \caption{ The persistence coefficient as a function of time for LCPs, HMFs, and IMFs. Dashed, dotted, dash-dotted and solid lines
 are for $^{40}$Ar+$^{45}$Sc, $^{86}$Kr+$^{93}$Nb, $^{86}$Kr+$^{197}$Au and $^{197}$Au+$^{197}$Au, respectively.}\label{fig1}
\end{figure}
\par
Let the number of pairs of nucleons in cluster C at time $t$ is
$\chi_{C}(t)$=0.5$\ast$$N_{C}$($N_{C}$-1). At the time $\Delta t$
later, it is possible that some of the nucleons belonging to
cluster C have left the cluster and are the part of another
cluster or are set free or others may have entered the cluster.
Now, let $N_{C_{D}}$ be the number of nucleons that have been in
the cluster C at time $t$ and are at $t+\Delta t$ in cluster D. We
define
\begin{equation}
\Phi_{C}(t+\Delta t) = \sum_{D}0.5\ast N_{C_{D}}(N_{C_{D}}-1).
\end{equation}
Here the sum runs over all fragments D present at time $t+\Delta
t$.
\par
The persistence coefficient of cluster C can be defined as
\cite{puri00,puri02}:
\begin{equation}
P_{C}(t+\frac{\Delta t}{2}) =\Phi_{C}(t+\Delta t)/\chi_{C}(t).
\end{equation}
The persistence coefficient averaged over an ensemble of fragments
is defined as:
\begin{equation}
\langle P(t+\frac{\Delta t}{2}) \rangle =
\frac{1}{N_{f_{t}}}\sum_{C}P_{C}(t+\frac{\Delta t}{2}),
\end{equation}
where N$_{f_{t}}$ is the number of  fragments present at time $t$
in a single simulation. The quantity is then averaged over a large
number of QMD simulations. The stability of a fragment between two
consecutive time steps can be measured through persistence
coefficient. If fragment does not emit a nucleon between two time
steps, the persistence coefficient is one. On the other hand, if
fragment disintegrates completely, the persistence coefficient
will be zero. If we remove one nucleon from a fragment C, the
persistence coefficient is P$_{C}$($t+\Delta t/2$) =
(N$_{C}$-2)/N$_{C}$ i.e., 0.333 for N$_{C}$ = 3 and 0.8 for
N$_{C}$ = 10. For example for mass 10, when one nucleon is emitted
we have two entities at later time step consisting of free nucleon
and fragment with mass 9. The P$_{C}$($t+\Delta t/2$) is the
contribution from all such entities existing at later times. It,
then, measures the tendency of the members of given cluster to
remain together. In fig. 1, we display the persistence coefficient
for various fragments i.e., light charged particles (LCPs)
(2$\leq$A$\leq$4), intermediate mass fragments (IMFs)
(5$\leq$A$\leq$44) as well as heavy mass fragments (HMFs)
(10$\leq$A$\leq$44). The various lines have been defined in the
caption. It is clear from fig. that the saturation value of
persistence coefficient is slightly higher in case of LCPs as
compared to heavier fragments. One can conclude that the final
fragments are formed after 130 fm/c when this coefficient is 0.8.
Before that time there is a strong exchange of nucleons between
the fragments. The number of medium and intermediate size
fragments increases because the largest fragment falls finally
into this mass bracket. The persistent coefficient reaches its
asymptotic value later due to the interaction between fragments as
well as between fragments and free nucleons. Due to this
interaction nucleons are sometimes absorbed or emitted from the
fragments. This process changes the details but not the general
structure of the fragmentation pattern.
\begin{figure}[!t]
\centering
 \vskip -1cm
\includegraphics[angle=0,width=15cm]{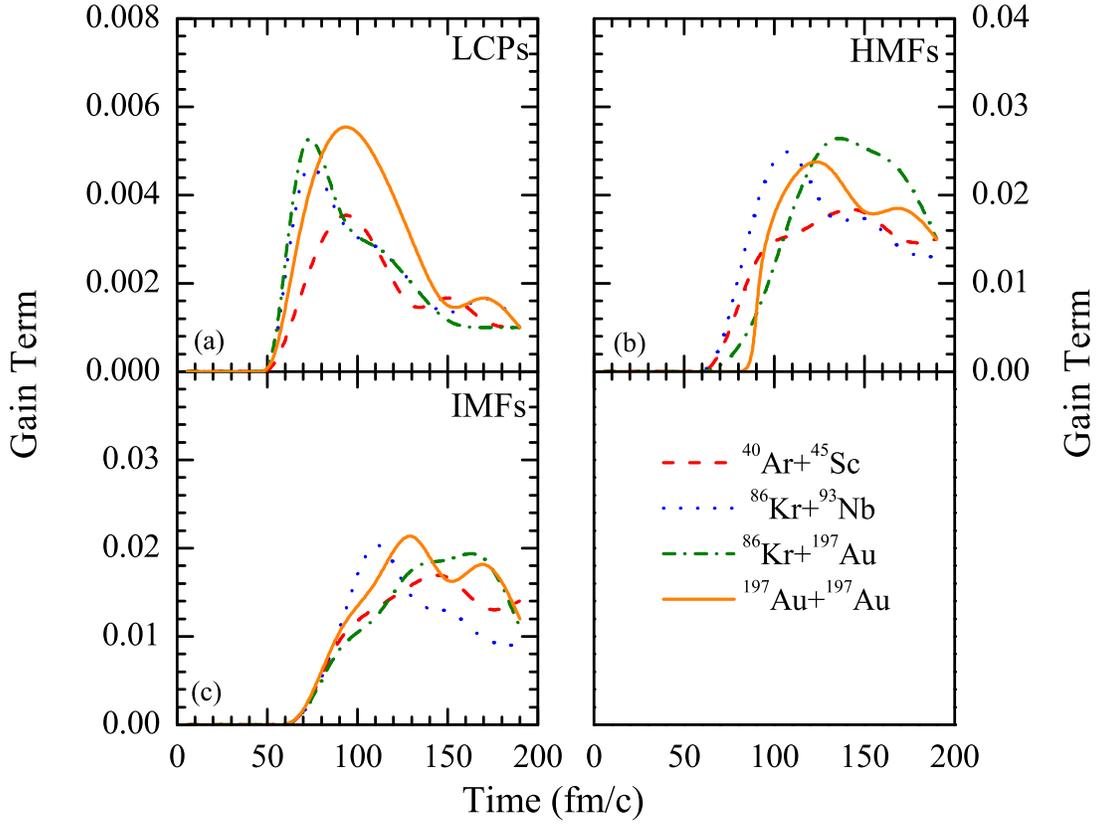}
 \vskip -6cm \caption{ The Gain term as a function of time for LCPs, HMFs, and IMFs. Lines have same meaning as in fig. 1.}\label{fig2}
\end{figure}
\par
The persistence coefficient tells about the stability of different
fragments between two successive time steps. But it does not
provide any information whether a fragment has swallowed some
nucleons or not. To check this, we use a quantity called "Gain"
\cite{puri02}. The Gain represents the percentage of nucleons that
a fragment has swallowed between two consecutive time steps. Let
N$^{f}_{\alpha}$ be the number of nucleons belong to a fragment
$\alpha$ at time $t$. Let N$^{f}_{\alpha \beta}$ be the number of
nucleons which were in cluster $\alpha$ at time $t$ and are in
cluster $\beta$ at time $t+\Delta t$. The Gain is defined as:
\begin{equation}
Gain(t+\Delta t/2)= \sum_{\alpha} \eta \times
\frac{\sum_{\beta}(N^{f}_{\beta}-N^{f}_{\alpha
\beta})}{N^{f}_{\alpha}};
\end{equation}

$\eta$ = 0.0, 0.5, and 1.0 if N$^{f}_{\alpha \beta}$ $<$
 0.5 N$^{f}_{\beta}$, N$^{f}_{\alpha \beta}$ = 0.5 N$^{f}_{\beta}$
and N$^{f}_{\alpha \beta}$ $>$ 0.5 N$^{f}_{\beta}$, respectively.
Naturally, a true Gain for a fragment $\alpha$ is only if its
nucleons constitute at least half of the mass of new fragment
$\beta$. The Gain term will tell us whether the interactions among
fragments have ceased to exist or not.
\begin{figure}[!t]
\centering
 \vskip -1cm
\includegraphics[angle=0,width=16cm]{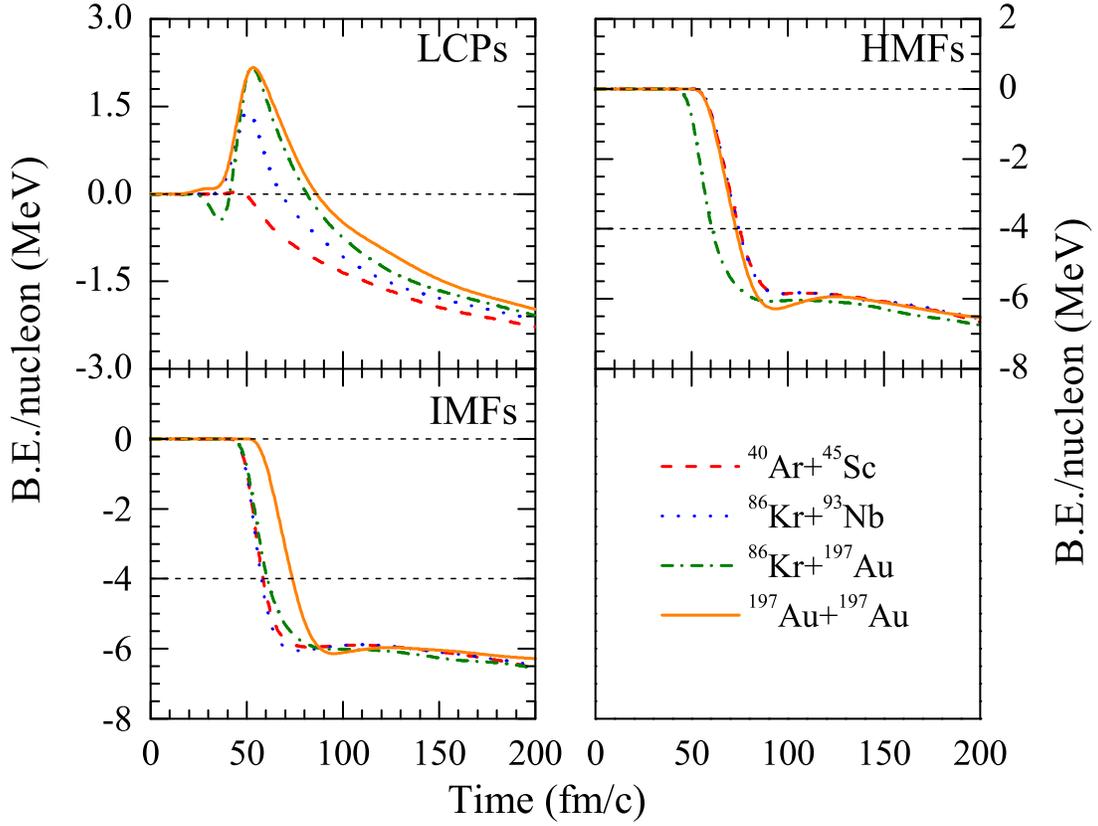}
 \vskip -6cm \caption{ The time evolution of binding energy per nucleon for LCPs, HMFs, and IMFs. Lines have same meaning as in fig. 1.}\label{fig3}
\end{figure}
\par
In fig. 2, we display the gain term for LCPs, HMFs, and IMFs. As
discussed earlier, the value of persistence coefficient is
slightly higher in case of LCPs. Therefore, gain term will be
smaller for LCPs as shown in fig. 2. As evident from fig., in case
of heavier systems the gain term has higher value because of the
large nucleon-nucleon interactions.

\par
To check the stability of fragments, we display in fig. 3, the
binding energy per nucleon as a function of time for LCPs, HMFs
and IMFs. We find that even at 200 fm/c, small fragments are still
not cold and take very long time to cool down. Whereas the heavy
fragments are properly bound having binding energy per nucleon
around -5 to -7 MeV.
\par
The quantities which are closely related to the degree of
thermalization are relative momentum $\langle K_{R}\rangle$ and
anisotropy ratio $\langle R_{a}\rangle$. The average relative
momentum of two colliding fermi spheres is defined as
\cite{puri94,khoaetal,khoa91}:
\begin{equation}
\langle K_{R} \rangle=\langle
|P_{P}(\textbf{r},t)-P_{T}(\textbf{r},t)|/\hbar \rangle,
\end{equation}
where
\begin{equation}
P_{k}(\textbf{r},t)=\frac{\sum_{j=1}^{A_{k}}P_{j}(t)\rho_{j}(\textbf{r},t)}{\rho_{k}(\textbf{r},t)}.
\end{equation}
Here P$_{j}$ and $\rho_{j}$ are the momentum and density
experienced by $j^{th}$ particle and $k$ stands for either target
or projectile and $\textbf{r}$ refers to a space point in central
sphere of 2 fm radius to which all calculations are made. The
$\langle K_{R} \rangle$ is an indicator of local equilibrium
because it depends on the local position \emph{r} .
\par
The second quantity is anisotropy ratio which is defined as
\cite{dhawan06,puri94,khoaetal,khoa91}:
\begin{equation}
\langle R_{a} \rangle=\frac{\sqrt{\langle p_{x}^{2}
\rangle}+\sqrt{\langle p_{y}^{2} \rangle}}{2\sqrt{\langle
p_{z}^{2} \rangle}}.
\end{equation}
The anisotropy ratio $\langle R_{a} \rangle$ is an indicator of
global equilibrium of the system because it represents the
equilibrium of the whole system and does not depend upon the local
positions. The full global equilibrium averaged over large number
of events will correspond to $\langle R_{a} \rangle$ = 1.
\begin{figure}[!t]
\centering \vskip -1cm
\includegraphics[angle=0,width=15cm]{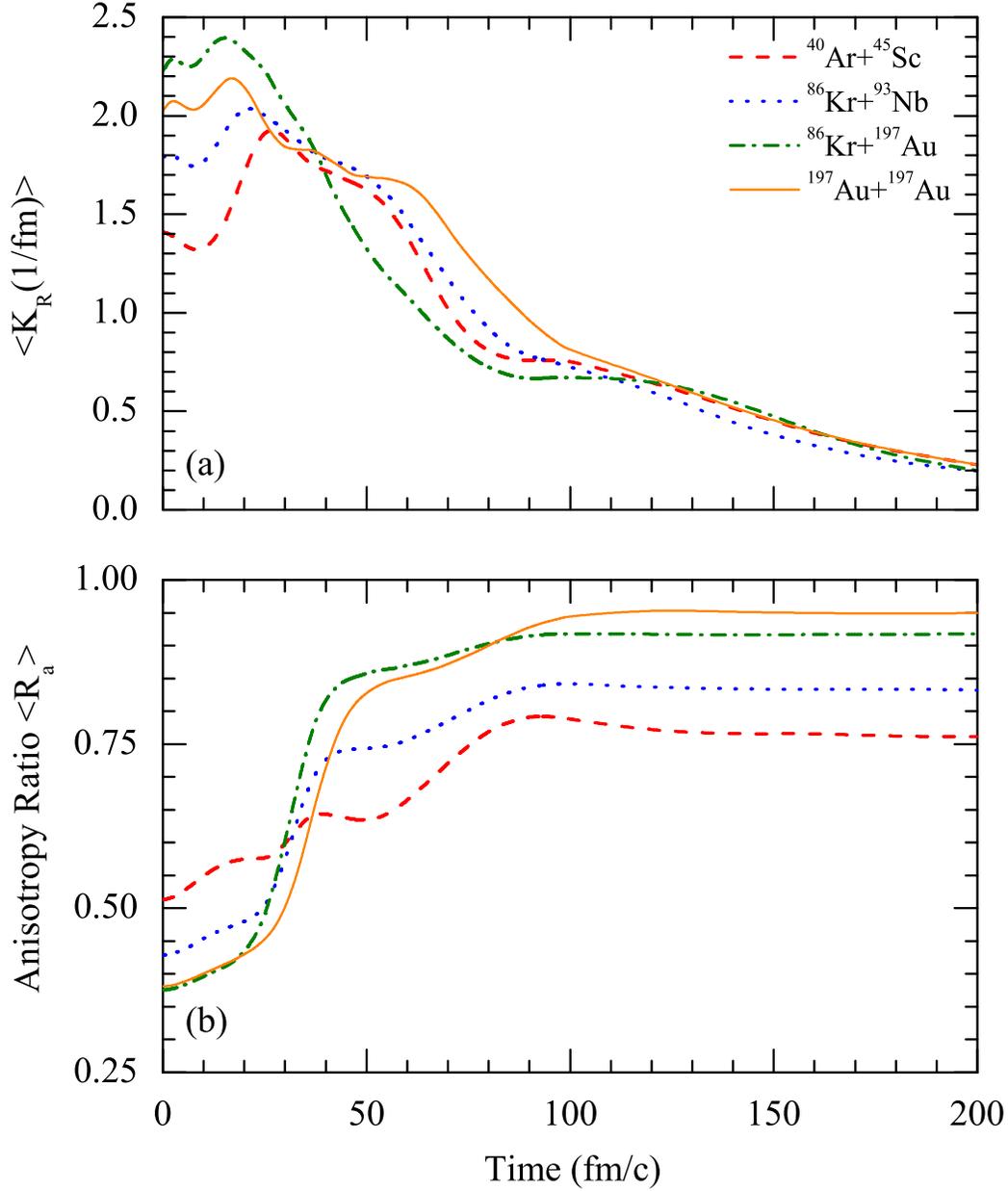}
\vskip -1.5cm \caption{The time evolution of (a) relative momentum
and (b) anisotropy ratio. Lines have same meaning as in fig.
1.}\label{fig4}
\end{figure}
\par
In figs. 4a and 4b, we display, respectively, $\langle K_{R}
\rangle$ and $\langle R_{a} \rangle$ ratio as a function of time
for different system masses. The initial value of relative
momentum increases whereas of the anisotropy ratio decreases with
mass of the system since E$_{c.m.}^{peak}$ increases with increase
in the system mass. It is interesting to see that the relative
momentum is large at the start of the reaction, and finally at the
end of the reaction, the value of $\langle K_{R} \rangle$ is
nearly zero. This means that at the end of the reaction, the local
equilibrium is nearly reached. However, the saturation time is
nearly the same throughout the mass range.  It is clear from the
fig. 4b anisotropy ratio changes to a greater extent during the
high density phase. Once the high density phase is over, no more
changes occur in thermalization. Interestingly, the heavier nuclei
are able to equilibrate more than the lighter nuclei. This is
because of the fact that the number of collisions per nucleon for
the $^{197}$Au+$^{197}$Au reaction is larger than for the
$^{40}$Ar+$^{45}$Sc reaction.
\begin{figure}[!t]
\centering
 \vskip -1cm
\includegraphics[angle=0,width=15cm]{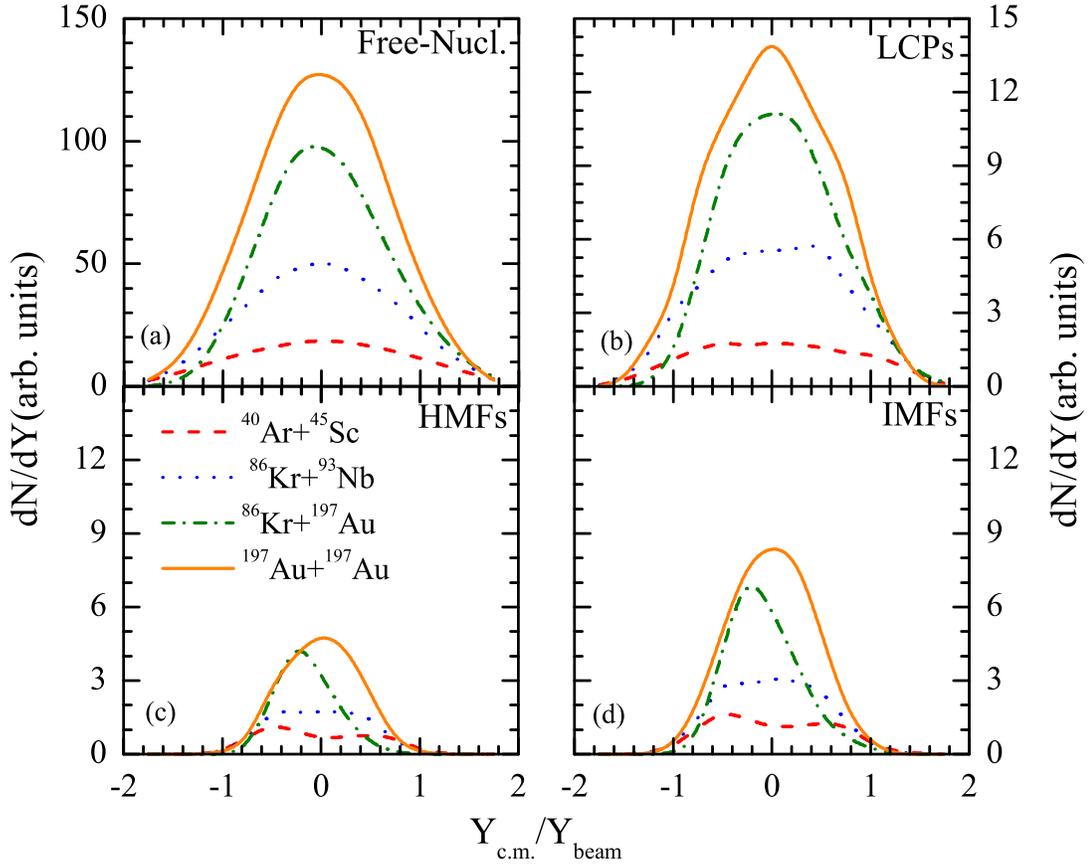}
 \vskip -6cm \caption{The rapidity distribution, dN/dY, as a function of reduced rapidity, Y$_{c.m.}$/Y$_{beam}$. Lines have same meaning as in fig. 1.}\label{fig5}
\end{figure}
\par
\begin{figure}[!t]
\centering
 \vskip -1cm
\includegraphics[angle=0,width=15cm]{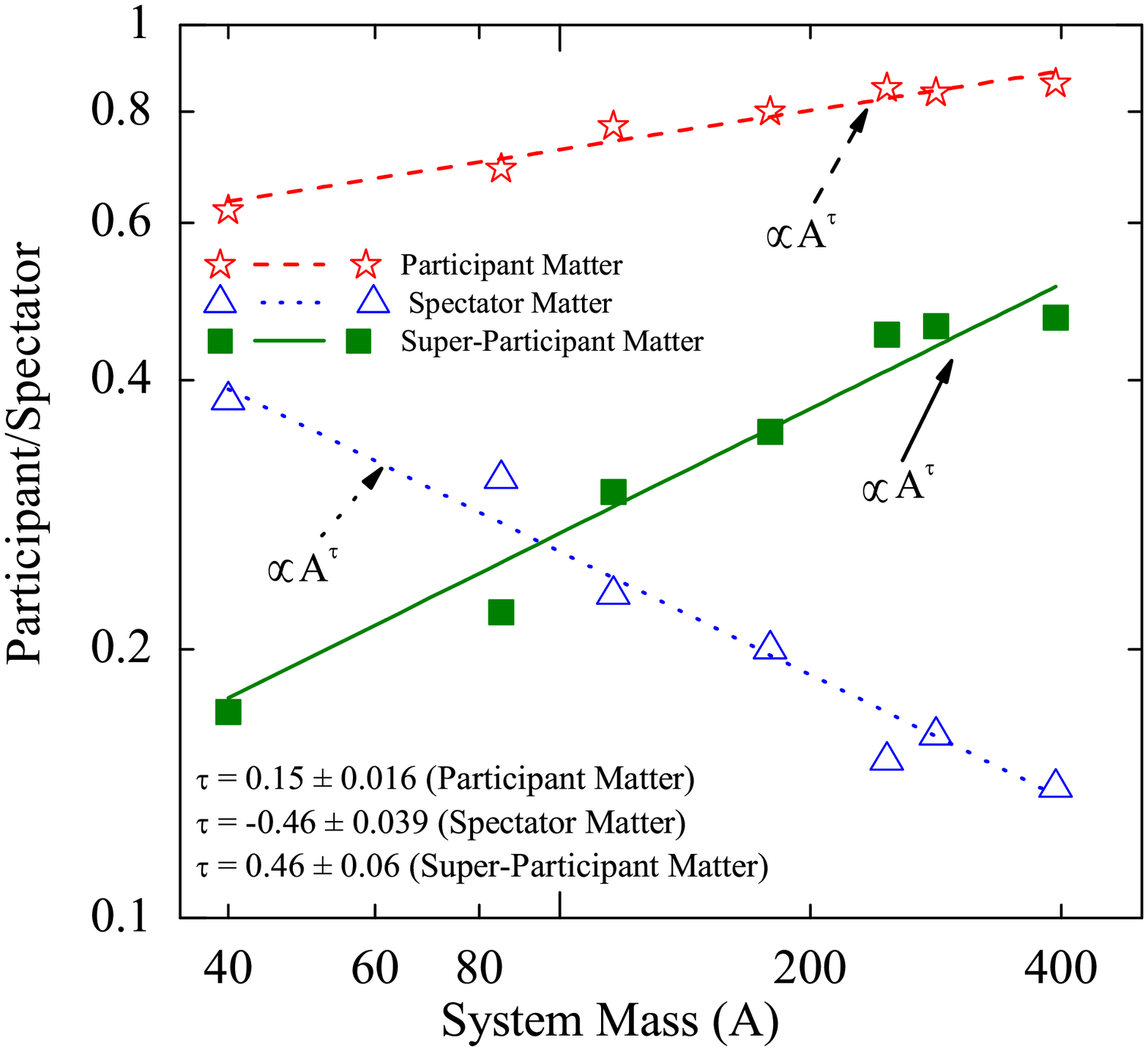}
 \vskip -6cm \caption{ The participant, spectator, and super-participant matter (defined in terms of nucleon-nucleon collisions) as a function of
 composite mass of the system. The dashed, dotted, and solid lines represent the the $\chi^{2}$ fits with power law A$^{\tau}$.}\label{fig6}
\end{figure}

\par
The rapidity distribution is also assumed to give information
about the degree of thermalization achieved in heavy-ion
reactions. The rapidity distribution of $i^{th}$ particle is
defined as \cite{dhawan06}:
\begin{equation}
Y(i)=\frac{1}{2}\ln\frac{\textbf{E}(i)+\textbf{p}_{z}(i)}{\textbf{E}(i)-\textbf{p}_{z}(i)},
\end{equation}
Here $\textbf{E}(i)$ and $\textbf{p}_{z}(i)$ are, respectively,
the total energy and longitudinal momentum of $i^{th}$ particle.
Naturally, for a complete equilibrium a single Gaussian shape peak
is expected. In fig. 5, we display the rapidity distribution of
free-nucleons, LCPs, HMFs as well as IMFs. Rapidity distribution
of all types of fragments indicate that heavier systems are better
thermalized as compared to lighter ones. For lighter nuclei, we
get relatively flat distribution. It is quite clear from the figs.
5(c) and 5(d) that very light nuclei ($^{40}$Ar+$^{45}$Sc) exhibit
a peak at target/projectile rapidity indicating a non-equilibrium
situation. Hence, for a complete stopping of incoming nuclei very
heavy systems are required. For HMFs and IMFs, in case of
$^{86}$Kr+$^{197}$Au, peak shifts towards left due to asymmetric
reaction. The effect is more pronounced for different kinds of
fragments as compared to free-nucleons.
\begin{figure}[!t]
\centering
 \vskip -1cm
\includegraphics[angle=0,width=15cm]{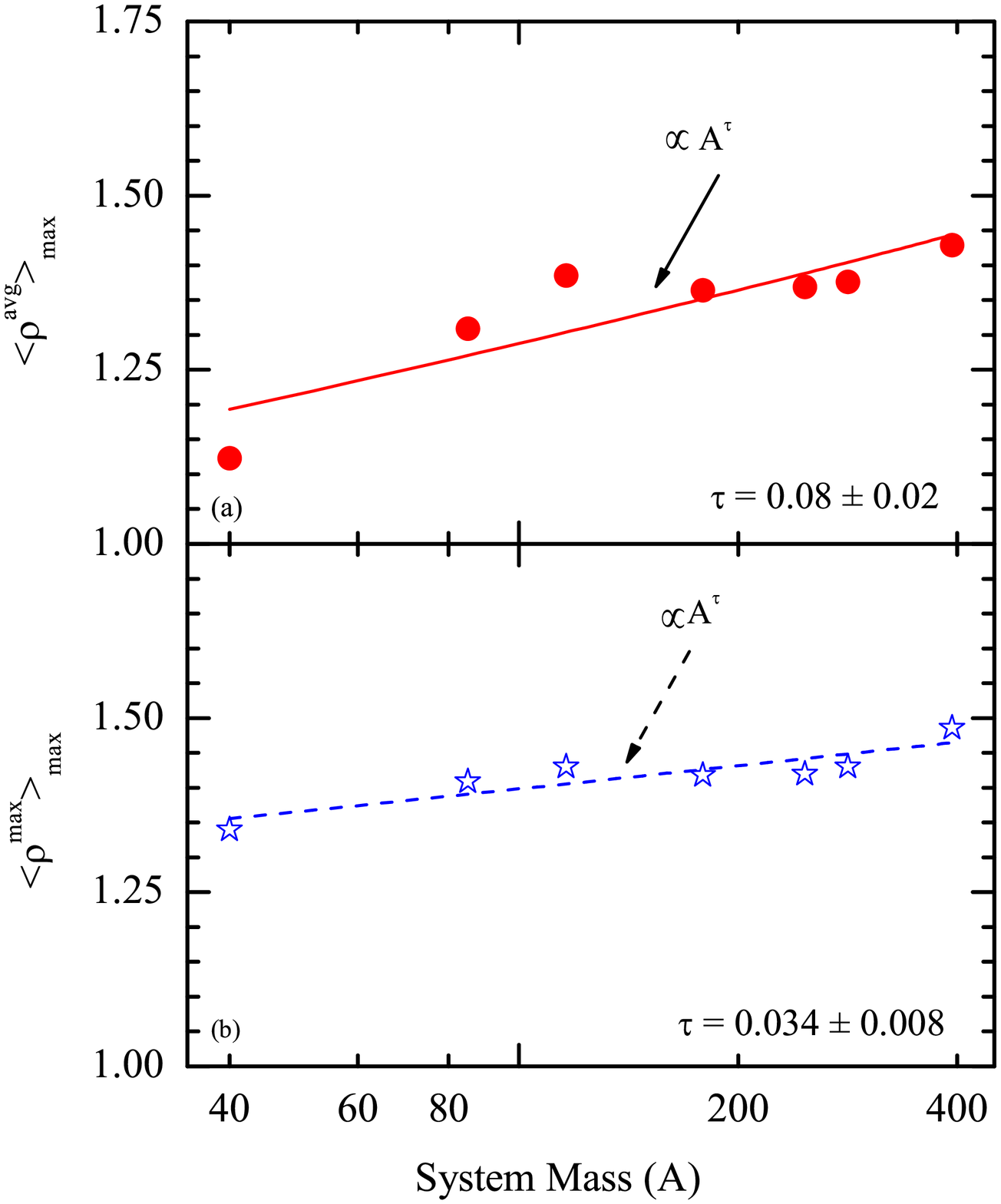}
 \vskip -1cm \caption{ The maximal value of the average density $\langle \rho^{avg} \rangle_{max}$ (upper part) and maximum density $\langle \rho^{max} \rangle_{max}$ (lower part)
 as a function of the composite mass of the system. The solid lines are the fits to the calculated results using A$^{\tau}$ obtained with $\chi^{2}$ minimization. The average is
 done over all space points on a sphere of 2 fm radius at center-of-mass. }\label{fig7}
\end{figure}
\par
In fig. 6, we display the participant, spectator, and
super-participant matter (obtained at 200 fm/c and defined in
terms of nucleon-nucleon collisions) as a function of the total
mass of the system. All nucleons that have experienced at least
one collision are termed as participant matter. The remaining
matter is counted as spectator matter. The nucleons having
experienced more than one collision are termed as
super-participant matter. As evident from the fig. 6, participant
matter shows a nearly mass independent behavior ( $\tau$ = 0.15 ±
0.016). The spectator and super-participant matter exhibit a power
law dependence with $\tau$ = -0.46 ± 0.039 for spectator matter
and $\tau$ = 0.46 ± 0.06 for super-participant matter.

\begin{figure}[!t]
\centering
 \vskip -1cm
\includegraphics[angle=0,width=15cm]{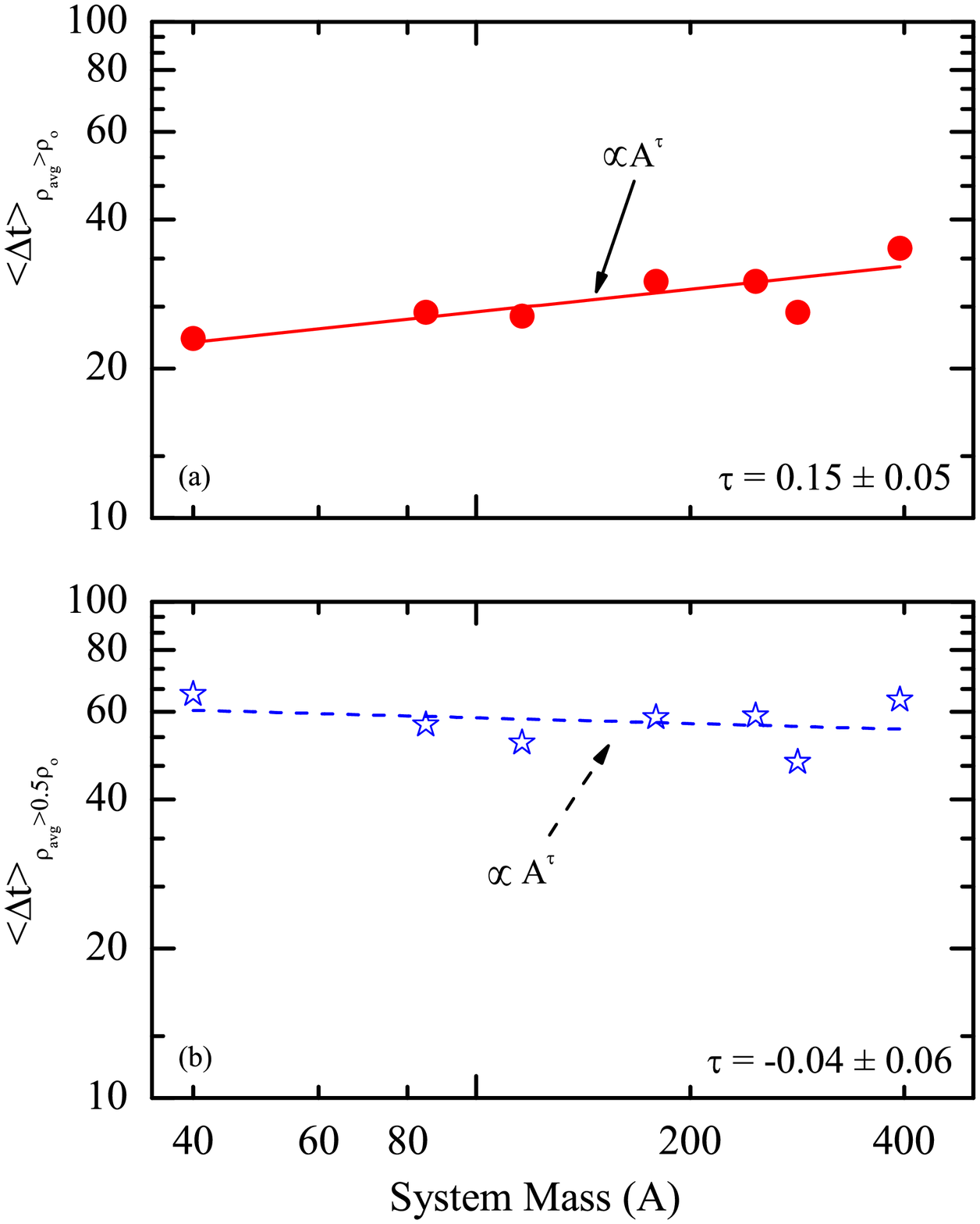}
 \vskip -1cm \caption{ The time zone for $\rho_{avg} \geq \rho_{o}$ (upper part) and for$\rho_{max} \geq \rho_{o}$ (lower part)
 as a function of composite mass of the system. The solid lines represent the the $\chi^{2}$ fits with power law A$^{\tau}$.}\label{fig8}
\end{figure}
\par
In fig. 7, we display the system size dependence of the maximal
value of average density $\langle \rho^{avg} \rangle$  (solid
circles) and maximum density $\langle \rho^{max} \rangle$ (solid
stars). Lines represent the power law fitting ($\propto$
A$^{\tau}$). The maximal values of $\langle \rho^{avg} \rangle$
and $\langle \rho^{max} \rangle$ follow a power law ($\propto$
A$^{\tau}$) with $\tau$ being 0.08$\pm$0.02 for the average
density $\langle \rho^{avg} \rangle$ and 0.034$\pm$0.008 for
maximum density $\langle \rho^{max} \rangle$ i.e., a slight
increase in density occurs with increase in the size of the
system. This is because, E$^{peak}_{c.m.}$ increases with the size
of the system.
\begin{figure}[!t]
\centering
 \vskip -1cm
\includegraphics[angle=0,width=15cm]{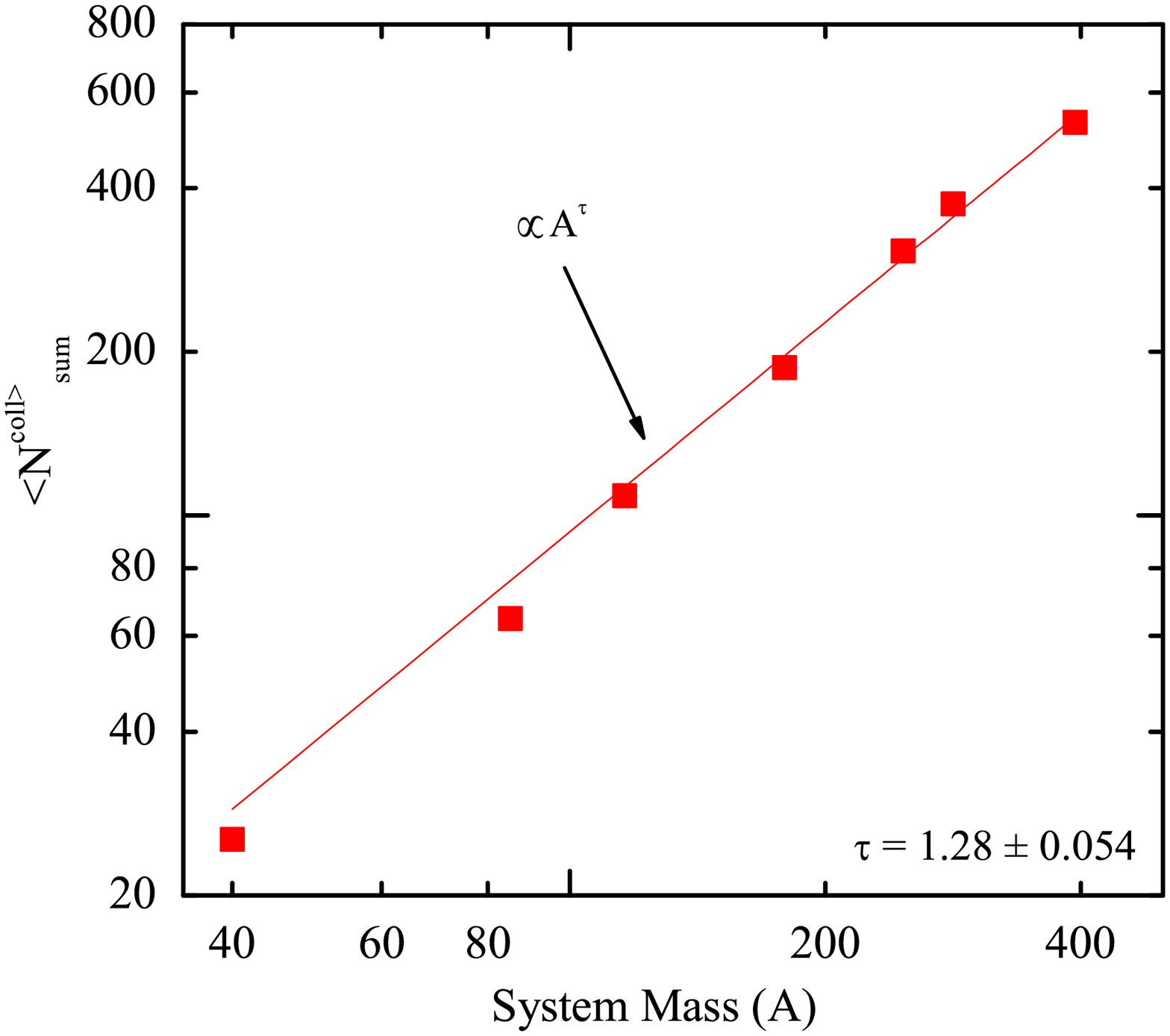}
 \vskip -5cm \caption{ The total number of allowed collisions versus composite mass of the system.
 The solid line represents the $\chi^{2}$ fits with power law A$^{\tau}$.}\label{fig9}
\end{figure}
 \par
Apart from the maximal quantities, another interesting quantity is
the dense zone at the peak energy. In fig. 8, we display the time
interval for which $\rho_{avg}\geq \rho_{o}$ (solid circles) and
$\rho_{avg}\geq \rho_{o}/2$ (open stars). Again both quantities
follow a power law behavior with $\tau$ = 0.15$\pm$0.05 and
$\tau$= -0.04$\pm$0.06, respectively, for $\rho_{avg}\geq
\rho_{o}$ and $\rho_{avg}\geq \rho_{o}$/2. This indicates that the
time duration for which $\rho_{avg}$ is greater than the normal
nuclear matter density increases with the mass of the system.

\par
The system size dependence of the (allowed) nucleon-nucleon
collisions (solid squares) is displayed in fig. 9. The results are
displayed at 200 fm/c where the matter is diluted and well
separated. The nucleon-nucleon collisions increase with the system
size. This enhancement can be parametrized with a power law
proportional to A$^{\tau}$ with $\tau$ = 1.28$\pm$0.054. At fixed
incident energy nucleon-nucleon collisions should scale as A. This
has been tested by Sood and Puri \cite{sood04}. Here power factor
is greater than one since with increase in mass of the system
E$_{c.m.}^{peak}$ also increases.

\section{Summary}
In the present study, we have simulated the central reactions of
nearly symmetric, and asymmetric systems, for the energies at
which the maximum production of IMFs occurs (E$_{c.m.}^{peak}$),
using QMD model. This study is carried out by using hard EOS along
with cugnon cross section and employing MSTB method for
clusterization. We have studied the various properties of
fragments. The stability of fragments is checked through
persistence coefficient, gain term and binding energy. We observed
that at 200 fm/c, small fragments are still not cold and they take
very long time to cool down whereas the heavy fragments are
properly bound. The information about the thermalization and
stopping in heavy-ion collisions is obtained via relative
momentum, anisotropy ratio, and rapidity distribution. We found
that for a complete stopping of incoming nuclei very heavy systems
are required. The mass dependence of various quantities (such as
average and maximum central density, collision dynamics as well as
the time zone for hot and dense nuclear matter) is also presented.
In all cases (i.e., average and maximum central density, collision
dynamics as well as the time zone for hot and dense nuclear
matter) a power law dependence is obtained.
\par
\section{Acknowledgements}
 This work has been supported by a grant from Department of
Science and Technology (DST), Government of India, India.
\par


\begin{thebibliography}{999}
\bibitem{goss97} P. B. Gossiaux and J. Aichelin, Phys. Rev. C {\bf
56}, 2109 (1997).
\bibitem{khoa92} R. K. Puri \emph{et al}., Nucl. Phys. A {\bf
575}, 733 (1994); Y. K. Vermani and R. K. Puri, Nucl. Phys. A
\textbf{847}, 243 2010.
\bibitem{khoaetal}D. T. Khoa \emph{et al}., Nucl. Phys. A {\bf
548}, 102 (1992).
\bibitem{aich91} J. Aichelin, Phys. Rep. {\bf 202}, 233 (1991).
\bibitem{tsang93} M. B. Tsang \emph{et al.}, Phys. Rev. Lett. {\bf
71}, 1502 (1993).
\bibitem{will97} C. Williams \emph{et al.}, Phys. Rev. C {\bf 55},
R2132 (1997).
\bibitem{dhawan06} J. K. Dhawan \emph{et al}., Phys. Rev. C {\bf 74}, 057901 (2006).
\bibitem{sood06} A. D. Sood \emph{et al}., Int. J. Mod. Phys. E {\bf
15}, 899 (2006); Y. K. Vermani \emph{et al}., Phys. Rev. C {\bf
79}, 064613 (2009).
\bibitem{puri94} R. K. Puri \emph{et al}., J. Phys. G: Nucl. Part. Phys. \textbf{20}, 1817 (1994).
\bibitem{sisan01} D. Sisan \emph{et al.}, Phys. Rev. C {\bf 63},
027602 (2001).
\bibitem{sukh10} Y. K. Vermani \emph{et al}., J. Phys. G: Nucl. Part. Phys. \textbf{36},
105103 (2009); S. Kaur and A. D. Sood, Phys. Rev. C  \textbf{82},
054611 (2010).
\bibitem{rkpuri} C. Hartnack \emph{et al}., Eur. Phys. J A {\bf 1}, 151 (1998);
S. W. Huang \emph{et al}., Phys. Lett. B \textbf{298}, 41 (1993);
 Y. K. Vermani and R. K. Puri, Europhys. Lett. {\bf
85}, 62001 (2009); Y. K. Vermani \emph{et al}., J. Phys. G: Nucl.
Part. Phys. {\bf 37}, 015105 (2010); S. Kumar \emph{et al}., Phys.
Rev. C {\bf 78}, 064602 (2008).
\bibitem{sood09} S. Kumar \emph{et al}.,  Phys. Rev. C {\bf 58}, 3494
(1998); A. D. Sood \emph{et al}., Phys. Rev. C {\bf 79}, 064618
(2009); S. Kumar \emph{et al}., Phys. Rev. C {\bf 81}, 014611
(2010); \emph{ibid} {\bf 81}, 014601 (2010); E. Lehmann \emph{et
al}., Prog. Part. Nucl. Phys. {\bf 30}, 219 (1993); E. Lehmann
\emph{et al}., Phys. Rev. C {\bf 51}, 2113 (1995).
\bibitem{jsingh00} J. Singh \emph{et al}., Phys. Rev. C \textbf{62}, 044617
(2000); G. Batko \emph{et al}., J. Phys. G: Nucl. Part. Phys. {\bf
20}, 461 (1994); S. W. Huang \emph{et al}., Prog. Part. Nucl.
Phys. {\bf 30}, 105 (1993); C. Fuchs \emph{et al}., J. Phys. G:
Nucl. Part. Phys. {\bf 22}, 131 (1996); A. D. Sood \emph{et al}.,
Phys. Lett. B \textbf{594}, 260 (2004); \emph{ibid} Phys. Rev C
\textbf{73}, 067602 (2006); S. Gautam \emph{et al}., J Phys. G:
Nucl. Part. Phys. \textbf{37}, 085102 (2010).
\bibitem{kumarpuri98} S. Kumar \emph{et al}.,  Phys. Rev. C {\bf
58}, 1618 (1998); R. Chugh and R. K. Puri, Phys. Rev. C
\textbf{82}, 014603 (2010).
\bibitem{jsingh} S. Gautam and A. D. Sood, Phys. Rev. C \textbf{83}, 014603 (2011);
\emph{ibid}., C \textbf{83}, 034606 (2011); S. Goyal, Phys. Rev. C
{\bf 83}, 047601 (2011); \emph{ibid}., Nucl. Phys. A.\textbf{853},
164 (2011).
\bibitem{puri00} R. K. Puri and J. Aichelin, J. Comp. Phys. {\bf
162}, 245 (2000); R. K. Puri \emph{et al}., Phys. Rev. C {\bf 54},
R28 (1996); P. B. Gossiaux \emph{et al}., Nucl. Phys. A
\textbf{619}, 379 (1997).
\bibitem{puri02} R. K. Puri \emph{et al}., Pram. J. Phys. \textbf{59}, 19 (2002).
\bibitem{dutt10} I. Dutt \emph{et al}., Phys. Rev. C {\bf 81}, 047601
(2010); \emph{ibid} {\bf 81}, 044615 (2010); \emph{ibid} {\bf 81},
064609 (2010); \emph{ibid} {\bf 81}, 064608 (2010); R. Arora
\emph{et al}., Eur. Phys. J A {\bf 8}, 103 (2000); R. K. Puri
\emph{et al}., Phys. Rev. C {\bf 43}, 315 (1991); \emph{ibid}
\textbf{45}, 1837 (1992); \emph{ibid} J. Phys. G: Nucl.Part Phys.
\textbf{18}, 903 (1992); \emph{ibid} Eur. Phys. J A \textbf{3},
277 (1998); \emph{ibid}  Eur. Phys. J A \textbf{23}, 429 (2005).
\bibitem{khoa91} D. T. Khoa \emph{et al}., Nucl. Phys. A \textbf{529}, 363 (1991).
\bibitem{sood04} A. D. Sood \emph{et al}., Phys. Rev. C {\bf 70},
034611 (2004).
\end{thebibliography}
\end{document}